\documentclass[journal,12pt,onecolumn,draftclsnofoot,]{IEEEtran}

\usepackage[margin=1.5in]{geometry}

\usepackage{graphicx}
\usepackage[utf8]{inputenc}
\usepackage{etoolbox}
\usepackage[hyphens]{url}
\usepackage{hyperref}
\PassOptionsToPackage{bookmarks=false}{hyperref}
\usepackage{color}
\usepackage[table,xcdraw]{xcolor}
\usepackage{gensymb}

\begin{document}

\title{Hybrid Urban Navigation for Smart Cities}

\author{Oisín Moran*, Robert Gilmore*, Rodrigo Ord\'o\~nez-Hurtado and Robert Shorten %
\thanks{This work was partially supported by Science Foundation Ireland grant 11/PI/1177.}
\thanks{*Joint first authors.}
\thanks{O. Moran, R. Gilmore, R. Ord\'{o}\~{n}ez-Hurtado and R. Shorten are with University College Dublin, School of Electrical, Electronic and Communications Engineering, Belfield, Dublin 4, Ireland.  Emails: oisin.moran@ucdconnect.ie, robert.gilmore@ucdconnect.ie, rodrigo.ordonez-hurtado@ucd.ie, robert.shorten@ucd.ie.}
}

\maketitle

\begin{abstract}

This paper proposes a design for a hybrid, city-wide urban navigation system for moving agents demanding dedicated assistance. The hybrid system combines GPS and vehicle-to-vehicle communication from an ad-hoc network of parked cars, and RFID from fixed infrastructure --such as smart traffic lights-- to enable a safely navigable city. Applications for such a system include high-speed drone navigation and directing visually impaired pedestrians. 

\end{abstract}


\section{Introduction}

Driven by increasing urban populations, ageing citizens, and an ever increasing need to make better use of existing resources, the smart city agenda has become one of the most defining research paradigms of recent times. At the heart of this research agenda is the need for services that require precise localisation and rapid navigation. Examples of such services include drones to assist emergency response units, and providing assistance to citizens with special needs. Thus, it is very likely that locating, controlling and feedback will be key enablers for a plethora of future city services.

Despite the obvious need for such services, their realisation is not without significant challenges. Many services that we wish to develop require precise and rapid navigation (drones to support emergency response) but should function even in the partial absence of power or sensor coverage, and at the same time be non-invasive and respect the privacy concerns of modern society.

Current navigation systems rely heavily on Global Navigation Satellite Systems (GNSSs) such as GPS and GLONASS, despite the fact that in urban environments GNSSs' signals are in general inaccurate. A consequence of such inaccuracies is increased power consumption (drifting around desired paths or waypoints will result in longer times to successfully complete a journey). As a result, drone navigation in urban scenarios tends to be vision-based, which is seen by many as a major privacy concern. Thus the objective of this paper is to propose a viable alternative to vision based systems that addresses some of these drawbacks.

Combining multiple types of sensing could provide the accuracy needed for dedicated autonomous navigation in cities, this paper focuses on creating such a system.
As it would be virtually impossible (or at least impractical) to instrument an entire city with battery-powered devices using traditional infrastructure, we propose merging conventional and non-conventional approaches to create a system that is a hybrid of high and low power sensing devices, and a hybrid of high and low precision positioning technologies. We argue that this type of system is the key to enabling large-scale robust, accurate, and rapid localisation of moving agents within a city.
For this, we propose to use an Ultra Wide Band (UWB) based system for the dense coverage of high power sensing, and propose to use a passive (cheap and zero power on the agent side) Radio Frequency Identification (RFID) based system for the sparser coverage of low power {\em``last-mile''} sensing.

The proposed hybrid system of wireless devices could be created by installing elements in the current city environment by either taking advantage of current infrastructure or by adding new sensors to the fleet of vehicles involved in ad-hoc networks. The latter approach, following \cite{Ordonez-Hurtado2017} also enables the possibility of alternative monetization platforms for connected vehicles, with a car owner potentially receiving micro-payments to participate in the network. Providing such a network of collaborative technologies offers a broader coverage of an urban environment. Such a dense and expansive network provides the possibility of creating accurate, real-time sensory systems using multiple technologies to improve localisation, which results in the creation of new, safe and robust navigation opportunities. Using passive RFID tags, for example, has the benefit of ensuring an agent is not restricted by the battery life of a sensor. This in effect produces a ``Zero Power" consuming sensor as the reflected radio waves in well-designed applications are enough to give adequate localisation in the event of a loss of power, representing a great advantage over traditional applications. In addition to this, this paper also investigates the viability of using a new style of static but flexible infrastructure, namely ad-hoc networks of parked cars, and how the potential services they provide may be able to support a multi-technology sensory system for localisation and navigation.

In summary, we aim to make possible rapid urban navigation for specialised moving agents, such as drones for emergency response and priority delivery or fully autonomous vehicles, by the adoption of a hybrid tracking/localisation system which enables wider coverage than conventional systems and a contingency method in cases of power shortages on the agent side. The efficacy of the proposed system is demonstrated by means of two important use-cases: (i) rapid drone navigation in an urban environment; and (ii) the development of a smart traffic light system to assist citizens with special needs.
%

\section{Enabling Technologies}

This paper investigates a number of technologies that could collect the data required to localise and track moving agents for control purposes. The infrastructure in which such technologies could be implemented, and the motivation for such a system, derive in part from the idea of a fluid infrastructure -- a system that can provide complete coverage from permanently fixed and temporarily stationary network nodes (road side units and parked cars respectively). This system would provide the opportunity to instrument a city and localise an agent at virtually every point of an agent's traversal through an urban environment.

The road side units could be adorned with RFID technology to allow ``Zero Power'' sensing in critical locations, while UWB technology could be implemented as part of a new fluid infrastructure utilising parked cars. The system and technologies are described in depth in the following section.

\subsection{Parked Cars as Infrastructure}

Cars are parked for up to 95\% of their lifespan \cite{sommer2014} and practically serve no purpose in such a state other than the economic drain of occupying valuable space. However, as an underutilised network of dense sensors, these otherwise idle cars lend themselves to new, creative and useful applications. For example they have been proposed for a wide variety of dedicated tasks from finding missing keys to detecting gas leaks \cite{Zhuk2016}. Additionally, with an increasing number of sensors, greater connectivity, and a move toward the car becoming a service platform, these potential use cases will only increase.

Traditional infrastructure is generally expensive to set up and also to repair and maintain. Using cars as infrastructure effectively eliminates these costs, as the fleet is gradually refreshed. This system also has the extra benefit of upgrading itself over time in contrast to the stagnation or degradation associated with traditional infrastructure. The large number of cars in cities and their spatiotemporal stability (staying in the same place for a long time) will also prove to be of great benefit. The implementation in question here is the integration of UWB antenna nodes within vehicles to act as data collectors and processing units for accurate position estimation purposes.

\subsection{Ultra Wide Band Technology}

With the recent explosion in the use of unmanned aerial vehicles (UAV), we have witnessed the legal landscape change (often not rapidly enough) in response to the real and perceived threats of drones. With vision-based navigation approaches proving so popular, the potential for the erosion of privacy becomes a growing and valid concern, with the additional downside that the performance of vision-based approaches rapidly deteriorates under poor weather conditions.
For applications where the end goal is not necessarily itself a visual task (such as delivering a package), and where privacy issues may be of great concern, it would be helpful to have a precise and rapid approach to navigation that is not dependent on the availability of a camera or other vision sensors. Even in cases requiring visual feedback, such as a fire station wishing to ascertain the existence or severity of a fire, having a system capable of navigating to the intended location without the need for a trained operator would save precious time for the firefighters who would now only need to be present for the final stage.

Currently, the closest option available to a precise and rapid non-vision-based approach would be the use of GPS which is not without its own technical drawbacks. A number of factors such as satellite position, signal attenuation, and clocking errors cause typical GPS applications to experience positioning errors of up to 30\,m on average, while the advanced version DGPS still yields an average error of 5\,m to 10\,m in a variety of environments \cite{kuo2013}. These positioning errors, specially those from conventional GPS, cannot be neglected in comparison with the average dimension of a drone (tens of centimeters). As with any traditional GNSS system, the accuracy of GPS can also be greatly affected by atmospheric/environmental conditions, shadowing by buildings, and multipath propagation which is especially problematic in dense urban environments \cite{Ordonez-Hurtado2014,Ordonez-Hurtado2017}. While there exist different demonstrations of GPS-only (i.e. non-cooperative) approaches for drone navigation, such as \cite{santana2015}, these are all invariably done in favourable conditions such as in open fields with no tall buildings.
As any inaccuracies in position estimation convert directly to speed reductions for waypoint tracking tasks, any increase in accuracy is highly desirable for time-critical applications. For emergency applications such as the defibrillator drone designed by Alec Momont \cite{dronesforgood}, these time savings could make all the difference for a victim of an out-of-hospital cardiac arrest where every minute that passes without treatment (CPR, defibrillatory shock, or definitive care) decreases his or her survival rate by around 5.5\% \cite{larsen1993}.

Our proposed system will thus attempt to solve both challenging issues: navigation speed and privacy. It will do so by using stationary cars as anchor (reference) nodes in a cooperative positioning approach. The cars will average out measurements of their GPS location, thereby reducing the error in their estimation of their location, and will only be added to the network after they have been stationary for a sufficient period of time (decided by a suitable performance index). Involved cars and the drone are required to be equipped with UWB-based systems to accurately estimate the distances between them and cooperatively localize the drone. In this paper we use UWB as a means to emulate current V2V communication systems based on Dedicated Short Range Communications (DSRC), but a UWB module could be easily added to the growing sensory array found in cars. Thus, as there is no imaging equipment involved in this approach, then no privacy concerns arise.

However, UWB-based systems depend on the availability of local power, so there is a gap in the system when power shortages occur in the host agent (i.e. UWB-based system is turned off due to the risk of a low battery level). In addition, as the density of cars in a city varies over space and time, there also exists a possibility that the network of parked cars may not always be sufficient to support such a localisation system. From a business district which may become car-sparse at night, to a pedestrian crossing where there may be no parking spaces nearby, holes may arise in the otherwise connected system. A system without a fail-safe mechanism to allow, for example, a visually impaired person to cross a pedestrian crossing under any circumstance would be neither robust nor safe. We thus propose that a viable method to fill these gaps and thereby make the system more robust, must rely on ``Zero Power'' technology such as that found in RFID systems.

\subsection{RFID Technology}

RFID systems are now an ever present part of modern life (generally hidden from sight), with applications that range from collecting tolls and paying bills to gaining access to buildings \cite{RFID_History}. In general, the traditional way of using RFID system is purely for identification purposes; however, by creating solutions that make use of different parameters, we find they are capable of much more. One such parameter that we will investigate in particular is the Received Signal Strength Indicator (RSSI) of a signal.

For the case of urban navigation, we want to be able to distinguish the RSSI values from different tags to estimate the distances to them, and then to work towards the goal of implementing control systems based on that information. Despite the fact that the RSSI readings are generally noisy, some tactics can be employed to improve/augment the quality of the collected signals, and thus ensure that the information found is trustworthy and distinguishable enough to create a robustly controlled feedback system. Some relevant aspects when using RFID systems are the technical issues when working under Non Line of Sight (NLOS) scenarios, this is, when there are obstructions between the tag and the receiver, either a varying or stationary obstruction. This paper does not investigate how to alleviate the concern of moving obstacles such as other road users (e.g., cars or people); however, there is already other research in this area (which can be further incorporated into our proposed approach). This paper focuses on applications in which RFID technology will not be affected by NLOS environments. In general, the environment will have a great impact on the measured RSSI values, however, with a suitable initial setup this issue can be alleviated.

Deploying RFID antennas capable of producing RF signals with sufficient power to produce identifiable backscattered signals at critical locations (e.g., traffic lights, uncontrolled intersections) is where this technology most clearly shows its benefits. At these points, both RFID and UWB could be used collaboratively to provide more accurate estimations of an agent's location. However, significantly at these critical locations, RFID could be used alone to determine the location and orientation of an agent in the event of a contingency related to power availability on the agent's side. Passive RFID tags allow for ``Zero Power" sensing on the agent's side, which is of huge potential benefit as it alleviates the dangers of loss of power in areas where safety is of utmost concern (pedestrian direction along a pedestrian crossing) or at the end point of a journey (drone navigation).

\section{Proposed Implementation}

Using both of the above technologies, namely UWB and RFID, integrated with parked cars as infrastructure, brings the potential to  virtually cover a city with enough units/sensors to make it fully navigable using a cooperative approach, and with a contingency strategy for critical scenarios. A critical location is, in general, a location at which agents are most at risk, and where the risk of losing signal or battery power would have the greatest effect.
A pedestrian crossing is a good example of such a location. Crossing, already a dangerous activity, has been made even more dangerous with the advent of ``quiet cars" as more electric and hybrid cars take to the roads. Thus, reading the user's location using both UWB and RFID technologies in conjunction could direct a user safely across a road.

\begin{figure}[h]
    \centering
    \includegraphics[scale=0.26]{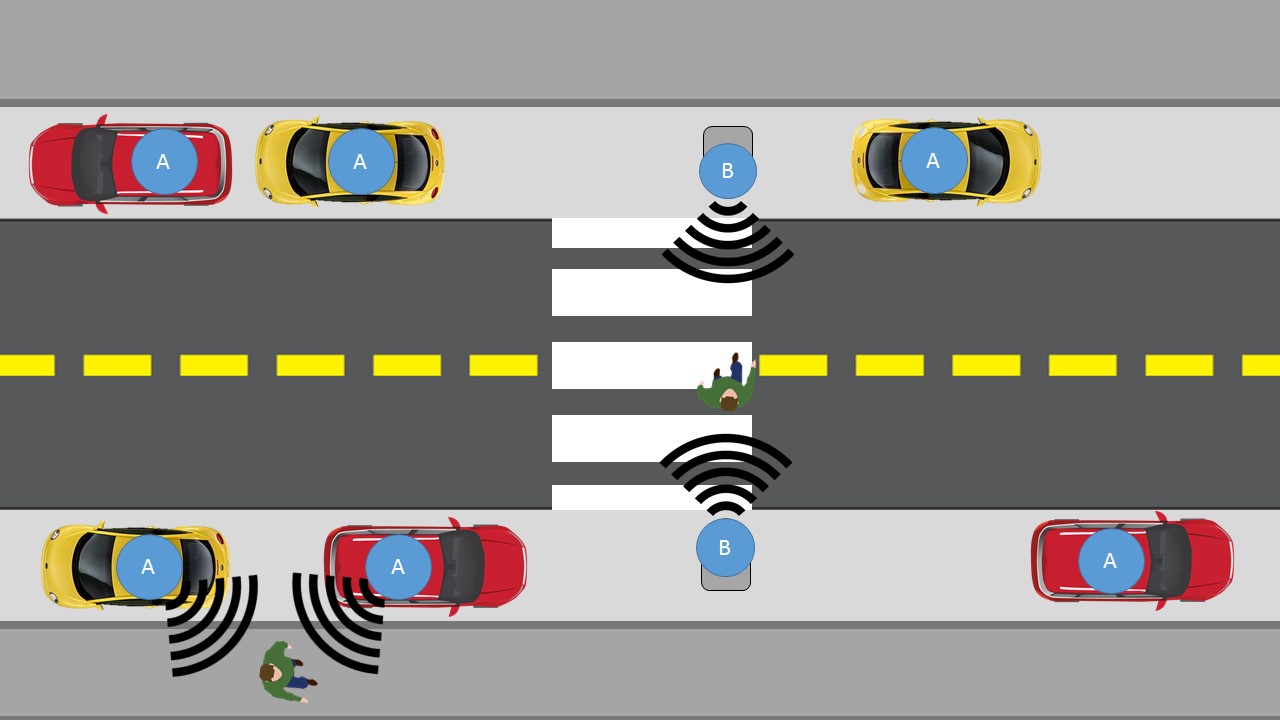}
    \caption{Potential infrastructure for urban navigation: pedestrians (moving agents), UWB-based systems (blue circles marked with A), RFID antennas (blue circles marked with B), and critical location (pedestrian crossing).}
    \label{fig:InfrastructureDiagram}
\end{figure}

\begin{figure}[h!]
    \centering
    \includegraphics[scale=0.33]{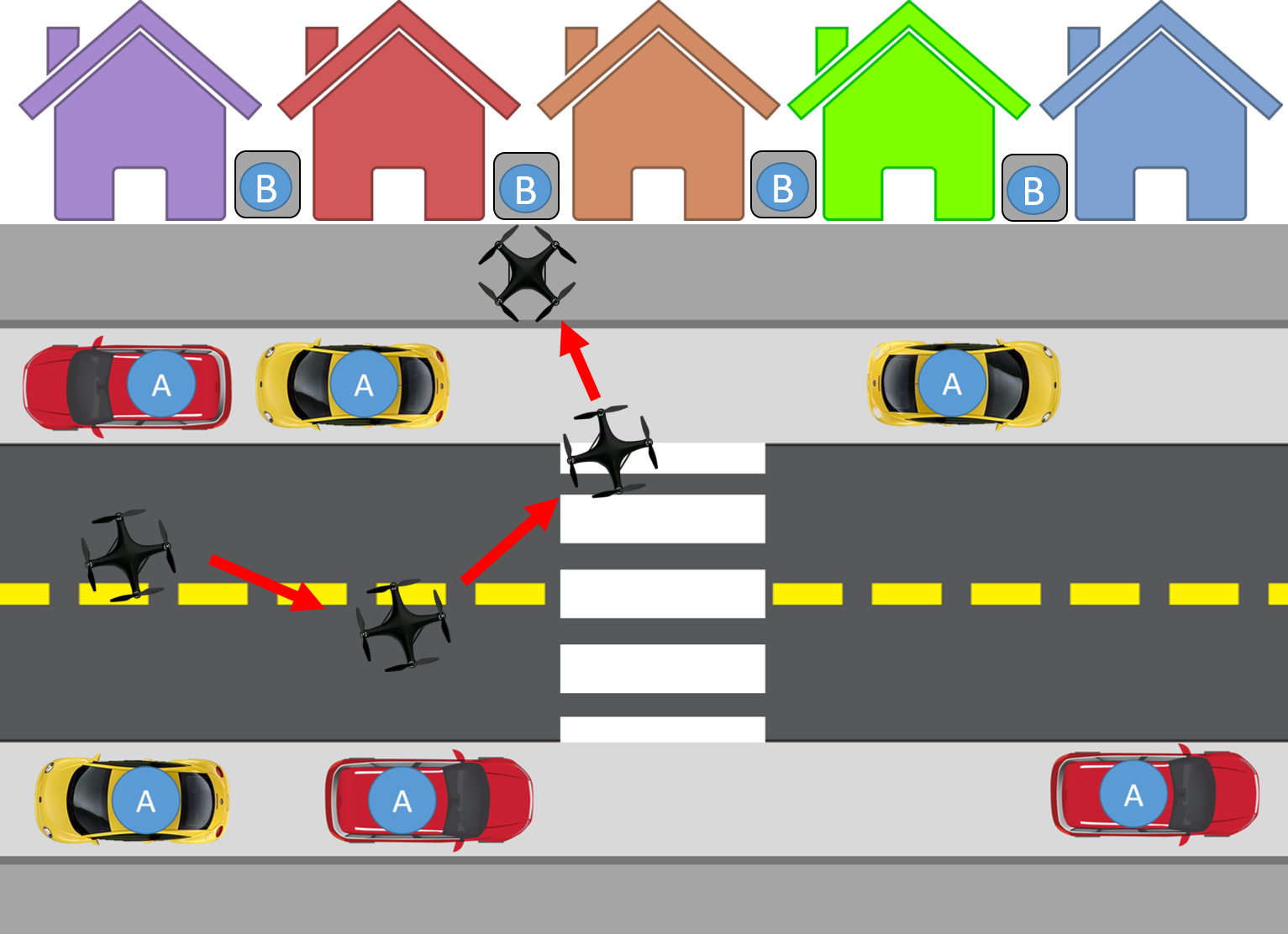}
    \caption{Potential infrastructure for drone delivery system: drone (moving agent), UWB-based systems (blue circles marked with A), and RFID antennas (blue circles marked with B).}
    \label{fig:DroneInfra}
\end{figure}

Figure \ref{fig:InfrastructureDiagram} illustrates a use-case scenario in which a visually impaired road user (a blind pedestrian) is directed using the UWB-based system for the largest section of their journey and then using both RFID and UWB, or RFID alone in critical zones. In the diagram, the cars are stationary (i.e. parked) and house UWB transceivers while the traffic light system would house the RFID antenna, thus adding functionality to existing infrastructure.
In the case of battery power loss, a visually impaired road user will be safe when on the pavement and can continue their journey with confidence as they can re-orient themselves with reference to a known point -- the edge of the pavement. However, at the critical and most dangerous point (the road crossing), no such reference is available. This is where the RFID system could provide feedback to the road user without need for any battery power on the agent's side.
Figure \ref{fig:DroneInfra} visualises a system that could be of interest to the implementation of package delivery using drones. Many companies are investigating the feasibility of drone delivery systems and the approaches discussed in this paper may be able to accurately control and position drones for this use. Again, the UWB positioning would be used for large-scale movement tracking, while RFID and UWB in conjunction could be used for more accurate positioning and orientation for docking, to ensure safe and timely delivery of products. Again, parked cars can be used as UWB anchors and RFID antennas could be installed like a traditional letterbox. RFID is useful in this use-case application for its potential even to control the orientation of the drone. Thus, in the case of a battery level at potentially critical level, then the UWB-based system could be turned off and the ``Zero Power" RFID-based sensing system would be hugely beneficial to complete an agent's journey successfully. 
 
\section{Experimental Results}

\subsection{Cooperative V2V navigation.}

\begin{figure}[h!]
\centering
\includegraphics[width=\linewidth]{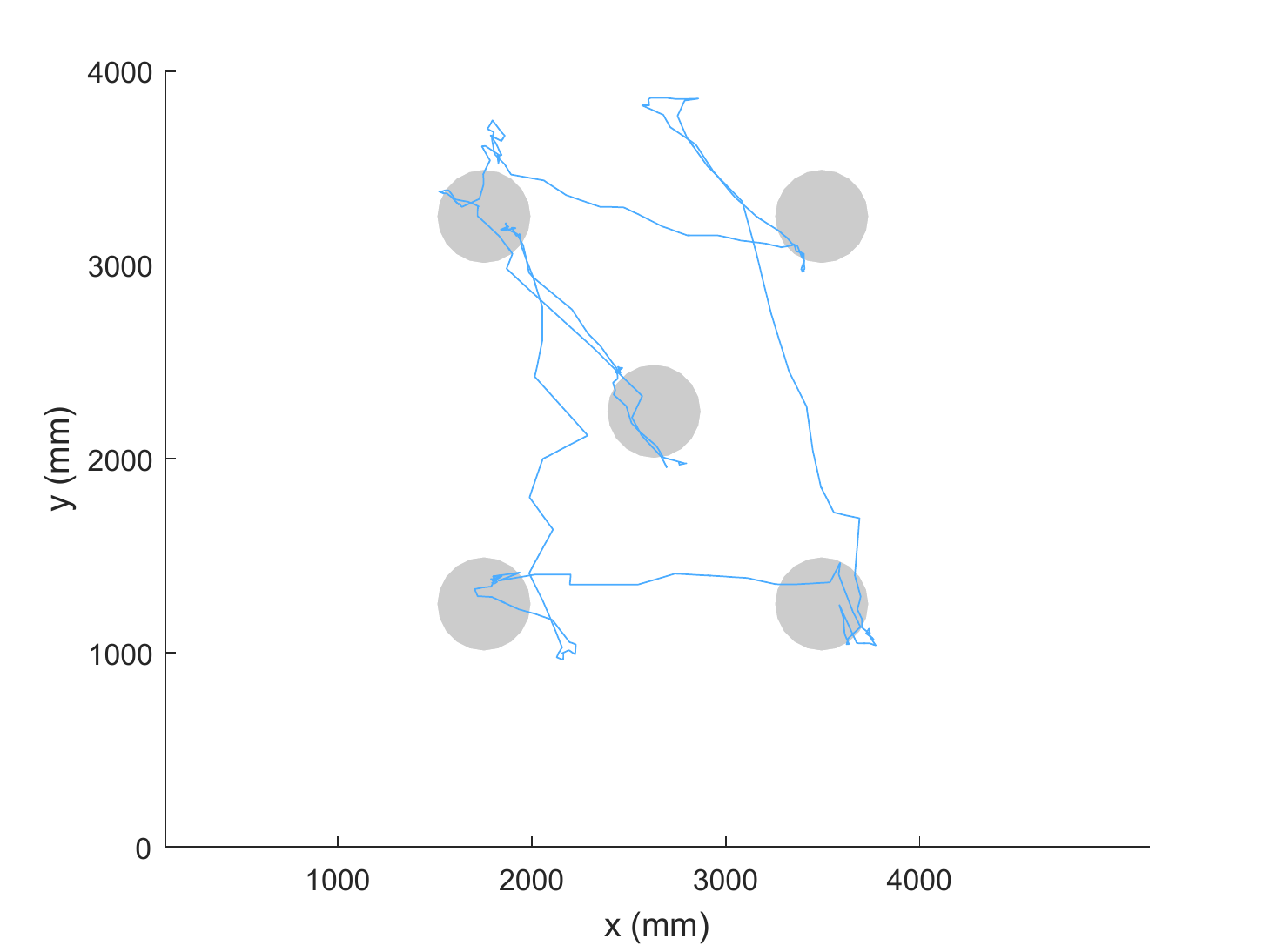}
\caption{Position of drone during indoor waypoint navigation task as estimated by the Pozyx system. Position of drone is shown in blue and waypoints of radius 25\,cm are shown in grey.}
\label{fig:drone_flight}
\end{figure}

Consistent indoor waypoint tracking has been successfully achieved using UWB-based Pozyx\footnote{{https://www.pozyx.io/}} positioning system.The drone was able to navigate to within a 25\,cm radius of each waypoint in a list, as demonstrated in Figure \ref{fig:drone_flight} and Movie A in Section \ref{Demonstrations}.

\begin{figure}[h!]
\centering
\includegraphics[width=\linewidth]{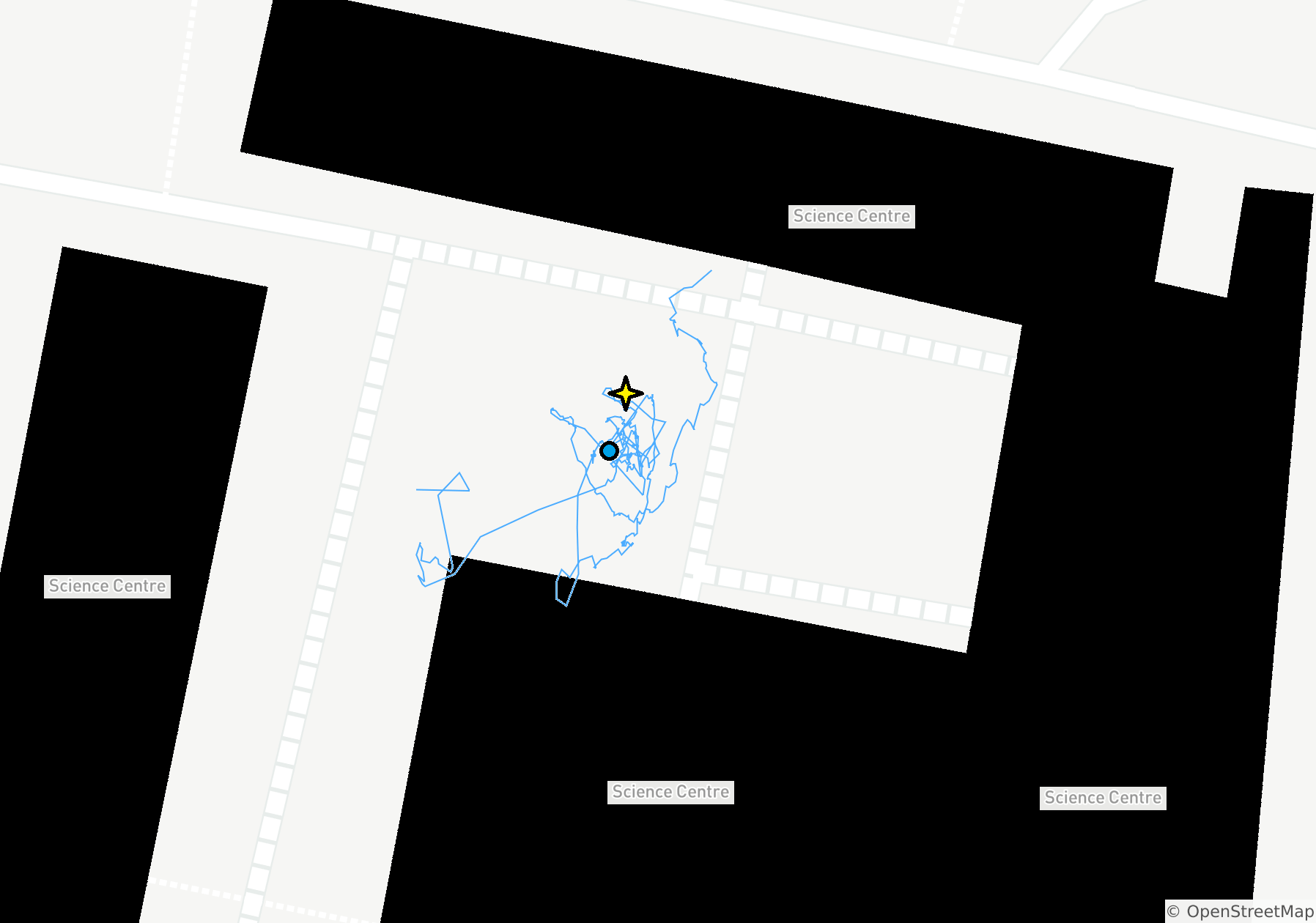}
\caption{GPS data (solid line) recorded every 10\,s over 4 hours, versus weighted average (blue circle) and true location (yellow star). Black areas correspond to tall buildings (UCD's Science Centre, Belfield, Dublin 4, Ireland).}
\label{fig:GPSreadings}
\end{figure}

Concerning tests in an urban-like scenario, we first established a GPS-only system as a benchmark for comparison reasons with respect to our cooperative proposed approach. The used location (periphery of UCD's Science Centre) was chosen to be representative of an urban environment (see Figure \ref{fig:GPSreadings}) and as such was very close to tall buildings. As the GPS readings at this location (solid line in Figure \ref{fig:GPSreadings})) contain positioning errors of high magnitude, the GPS-only system was unworkable in such an scenario, which provides a stronger justification to the use of our approach.
So far, in the steps toward a full proof of concept, we used the whole set of collected GPS data (4 hours of latitude, longitude, and HDOP readings, collected every 10\,s) to obtain an improved position estimation by means of a weighted average described in \cite{Ordonez-Hurtado2017}. Thus, every sample $i$ is weighted by the factor $\lambda_i$ given by
\[\lambda_i = \frac{HDOP_i^{-2}}{\sum_{j=1}^N HDOP_j^{-2}}\]
to finally obtain the average value represented with a blue circle in Figure \ref{fig:GPSreadings}, which is within a proximity of 2\,m from the real location (yellow star). With these results, our next step will be the validation of the waypoint tracking in the outdoor scenario using the hybrid (GPS+UWB) positioning system, where the relative positioning can be obtained with high accuracy using the UWB anchors, and then the absolute position is estimated by merging the information from the weighted average of GPS positions. Such a validation process is currently being prepared.

\subsection{RFID localisation using RSSI measurements}

\begin{figure}[h]
    \centering
    \includegraphics[width=\linewidth,trim={1.8cm 2cm 1.8cm 2cm}]{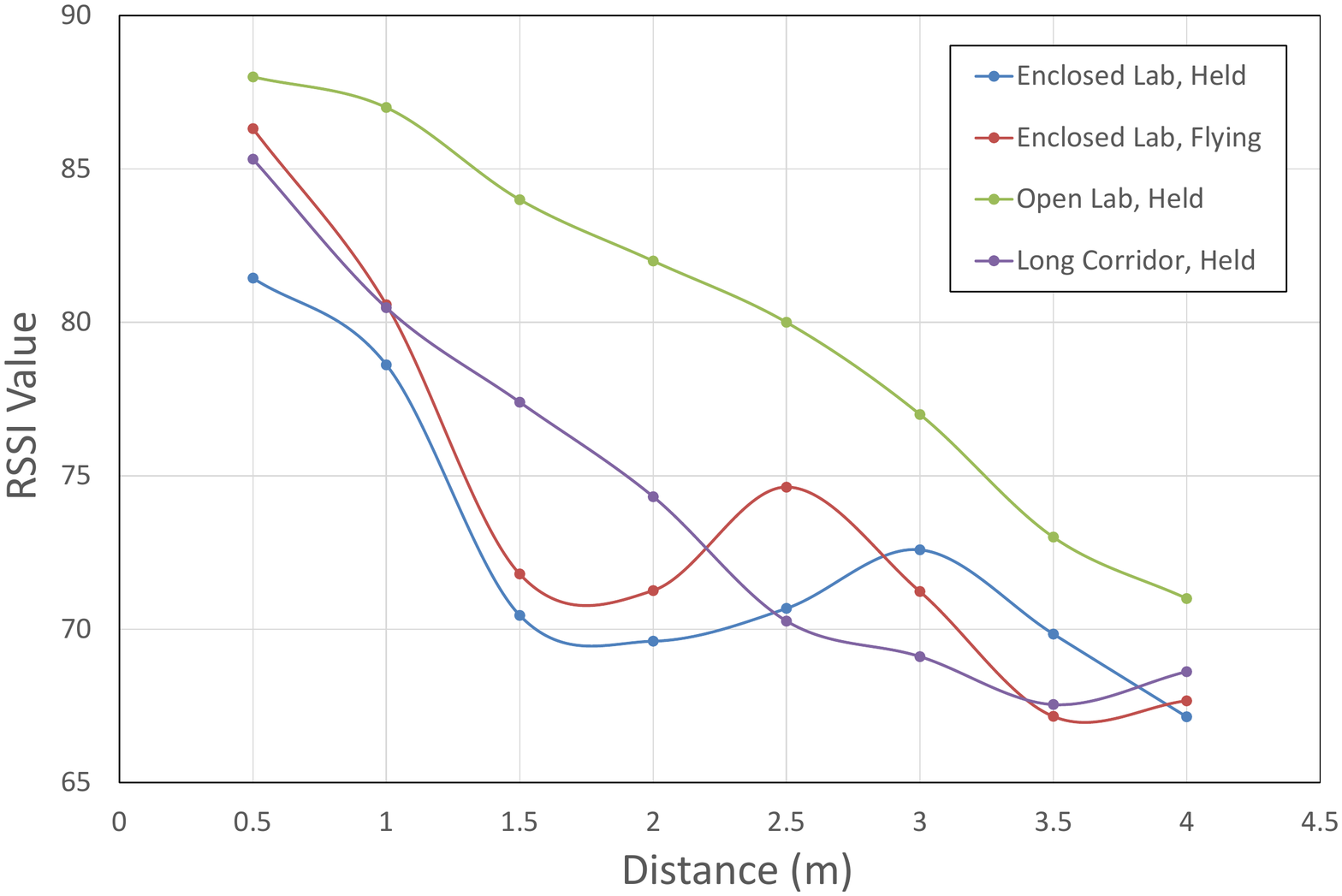}
    \caption{RSSI curves for a tag a placed on drone in different environments and operation modes. Each point correspond to the standard average of over 100 reading collected over 20\,s.}
    \label{Curve3}
\end{figure}

Characterising the relationship between RSSI  readings and distance/orientation of moving agents carrying RFID tags is of utmost importance; this is to provide evidence that the backscattered RF signals are reliable and accurate enough to be used in navigational control systems.
As mentioned previously, there are some limitations in the use of RFID technology, one of which being that the environment in which it is operated greatly impacts the readings obtained from tags. We have investigated this experimentally with a Parrot AR.Drone 2.0 covered by few tags, and the resulting curves displayed together in Figure \ref{Curve3}.
The differences in the returned RSSI values can be taken into account before making a decision in any control system, or the system could be constructed in a known location in which the environment is suitable. Something to note is that applications in a smart city context will be more likely to involve outdoors environment where backscattering and multipath effects of RF waves will not be of much concern. The locations of such systems can be chosen to minimize dynamic changes in the environment, such as elevated locations including the top of traffic lights and lamp posts.

Figure \ref{Curve3} shows the standard average over 20\,s of RSSI readings per location, resulting in over 100 readings per location. This test was undertaken in multiple environments (enclosed/open) and operation modes (flying/held) to gain insight on how the location of the controllable agent would affect the returned RSSI values. For the enclosed laboratory, it is seen that for the first 2\,m, the curve can be characterised as a straight line, whereas at further distances it would be more difficult to determine where the tag is located. There is a one-to-many relationship caused by multipath effects in such an environment. This would not be the case in an outdoor environment as proven by the straight line relationship seen in the more open and unobstructed laboratory. Furthermore, it is useful to note that the RSSI values are of larger magnitude in the open lab, which is likely due to the reduction in multipath reflected return signals that would reduce the average value over the experimental test period.

\subsubsection{Distance Estimation}

\begin{figure}[h]
    \centering
    \includegraphics[width=\linewidth,trim={1.8cm 2cm 1.8cm 2cm}]{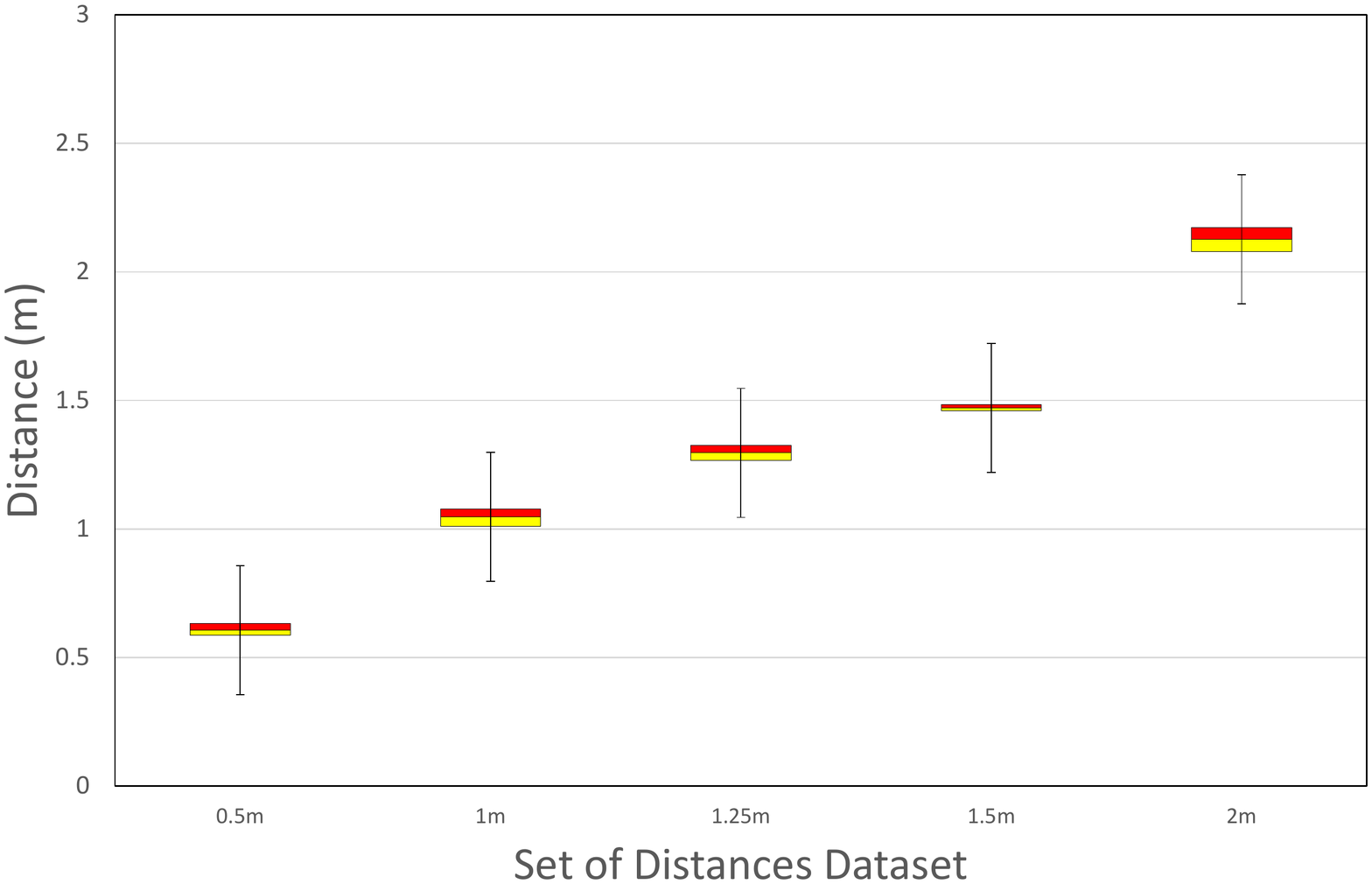}
    \caption{RSSI curve for tags attached to a drone at data points located in between 2 antennas, in an open laboratory. Each data point was interrogated for 10\,s, which resulted in over 80 values per data point.}
    \label{Curve2}
\end{figure}

The results displayed in Figure \ref{Curve2} are for distance estimation of a drone carrying multiple tags with respect to two RFID antennas. The antennas were situated at either end of a straight line, with the tags on the drone orientated perpendicularly with respect to the plane of the surface of both antennas.
It should be noted here that despite the fact that maximum/minimum readings have a large variation the inter-quartile ranges are relatively small and do not overlap. In this experiment the RFID tags were interrogated for 10\,s per data point which equated to no less than 80 returned RSSI values.

Now that we can characterise the relationship between RSSI and distance, a Java script was developed to record the RFID-based position estimations. The information shown in Table \ref{ErrorTable} reveals that the system is accurate to within 13\,cm, which is relevant enough for the dimensions of the test drone (51.7\,cm x 51.7\,cm).

\begin{table}[h]
\caption{RFID-based position estimation: averaged positioning error in recorded estimations}
\centering
\begin{tabular}{|>{\columncolor[HTML]{EFEFEF}}c |c|c|c|c|c|}
\hline
True Dist (m)         & 0.5     & 1       & 1.5     & 2      & 2.5     \\ \hline
Avg Recorded Dist (m) & 0.6023  & 1.0449  & 1.3011  & 1.4738 & 2.1338  \\ \hline
Error (m)                 & -0.1023 & -0.0449 & -0.0511 & 0.0262 & -0.1338 \\ \hline
S.Deviation (m)    & 0.0489  & 0.0503  & 0.0422  & 0.0212 & 0.0622  \\ \hline
\end{tabular}
\vspace{0.1cm}
\label{ErrorTable}
\end{table}

\begin{figure}[h]
    \centering
    \includegraphics[width=\linewidth,trim={1.8cm 2cm 1.8cm 2cm}]{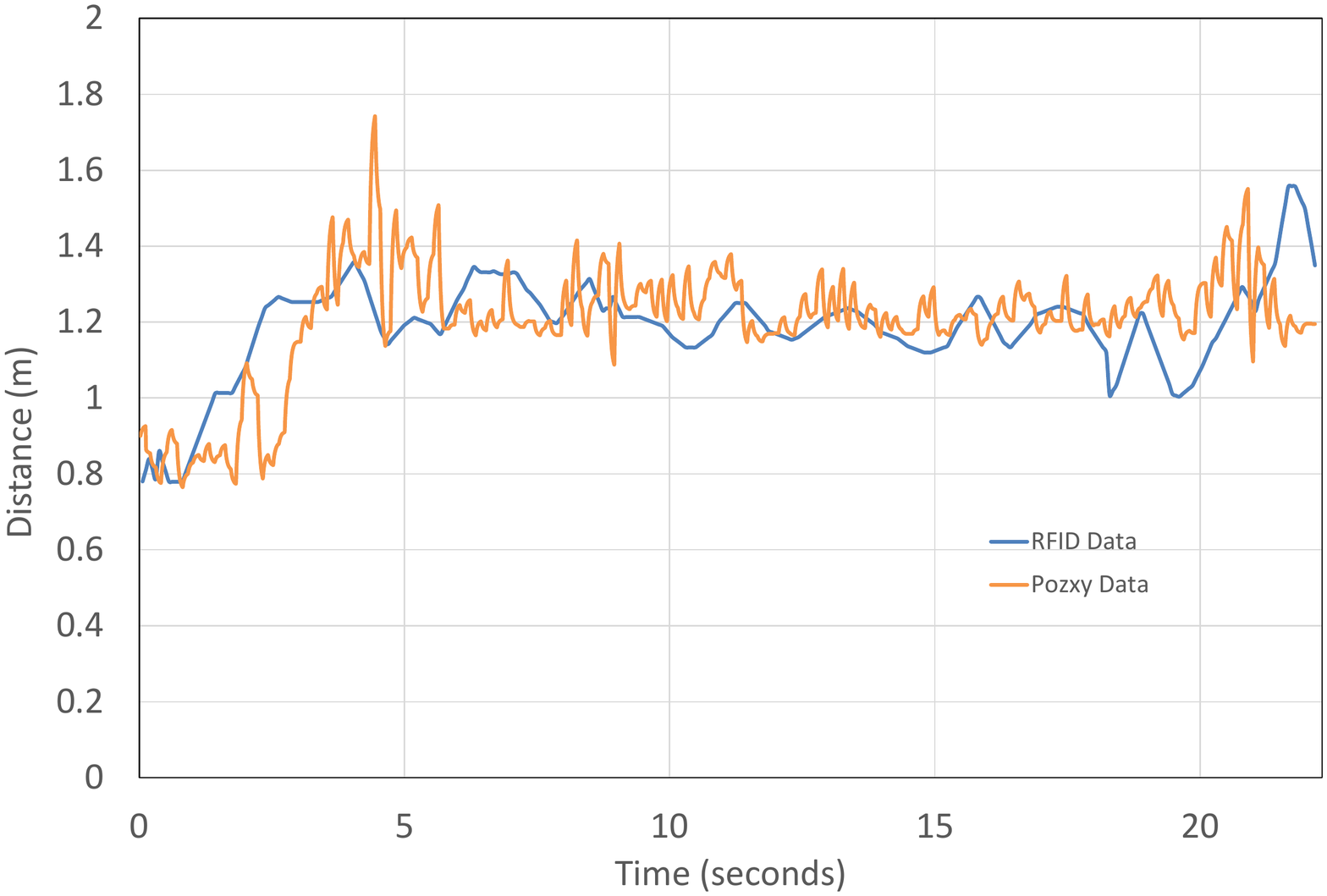}
    \caption{Location estimation from RFID system and Pozxy system: 0.8\,m to 1.25\,m.}
    \label{Path1}
\end{figure}

Figure \ref{Path1} displays the information recorded during a straight line flight test, in which a simple control system is implemented. A destination point was set and the drone placed by the user at a point away from the destination. In this case, the starting point was 0.8\,m from the reference antenna, and the destination was 1.25\,m from the reference antenna. The RFID system collected the RSSI values and converted them to a distance; this estimation using RFID is shown in Figure \ref{Path1}. For comparison, the approximate ground truth  was measured by the Pozyx system. A link to a video of this in operation can be found in Section \ref{Demonstrations}.

\subsubsection{Angle Estimation}

\begin{figure}[h]
    \centering
    \includegraphics[width=\linewidth,trim={1.6cm 2cm 2.1cm 2cm}]{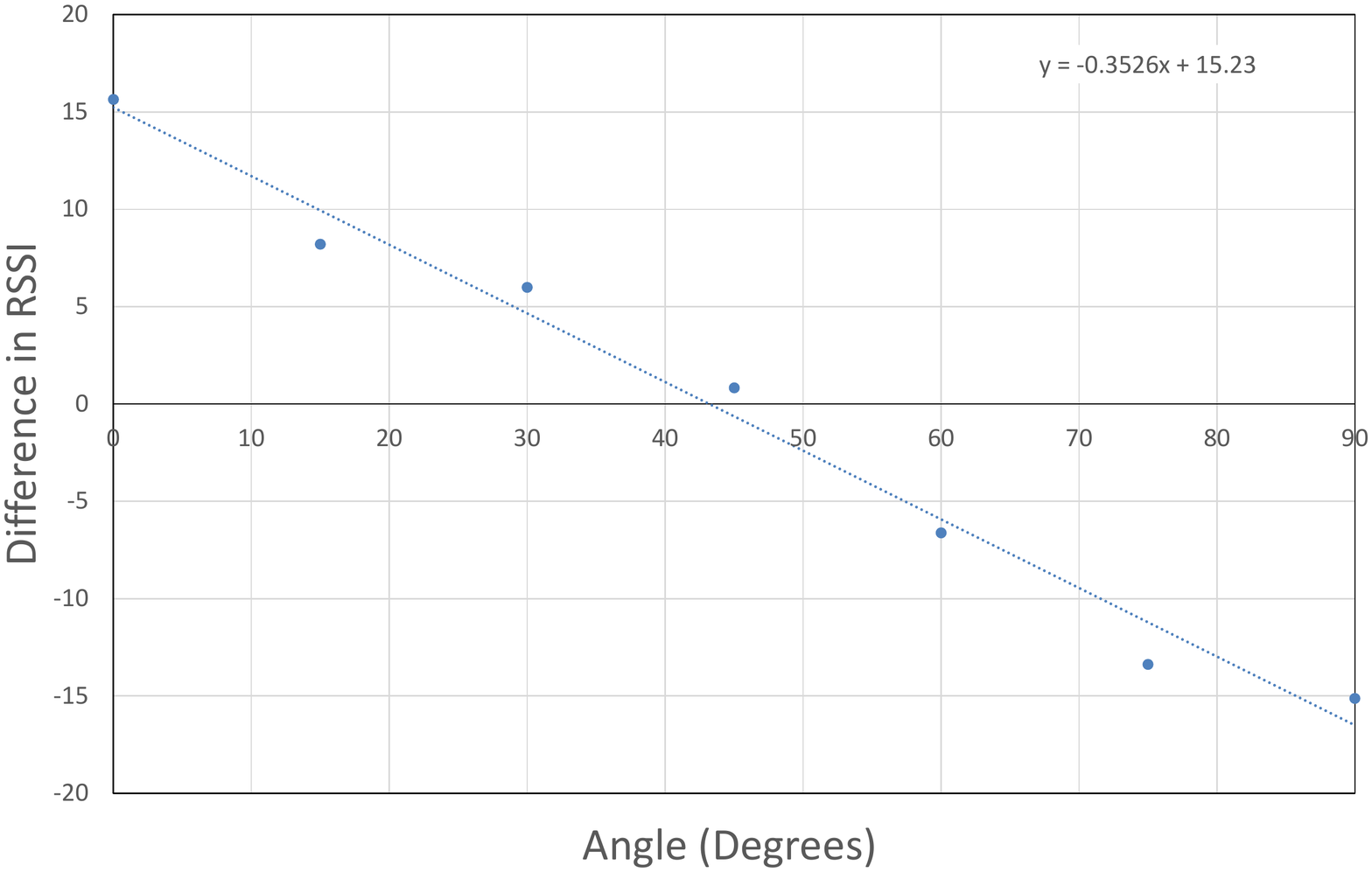}
    \caption{RSSI curve for the difference between the average value from 2 tags affixed to one side of the drone with 40\,cm of separation.}
    \label{anglegraph2}
\end{figure}

For many applications, the orientation of an agent is useful or even necessary. Investigations into how the comparison of multiple tags positioned (at known locations) on one agent has been undertaken here. Emphasis is placed on analysing the differences in returning RF signals, that when processed could track orientation in real-time. The results prove that not only distance can be accurately measured using RFID, but a relationship can be formed that converts the differences in RSSI to corresponding angles. An example of this data collected from 2 different tags on one side of the drone is shown in Figure \ref{anglegraph2}, in which an roughly linear relationship is observed. Furthermore, Table II shows that the error between the true angle and the recorded angle using RFID technology is on average less than 1.1 degrees over a time period of 5\,s. 

\begin{figure}[h]
    \centering
    \includegraphics[width=\linewidth,trim={2.5cm 2cm 2cm 2cm}]{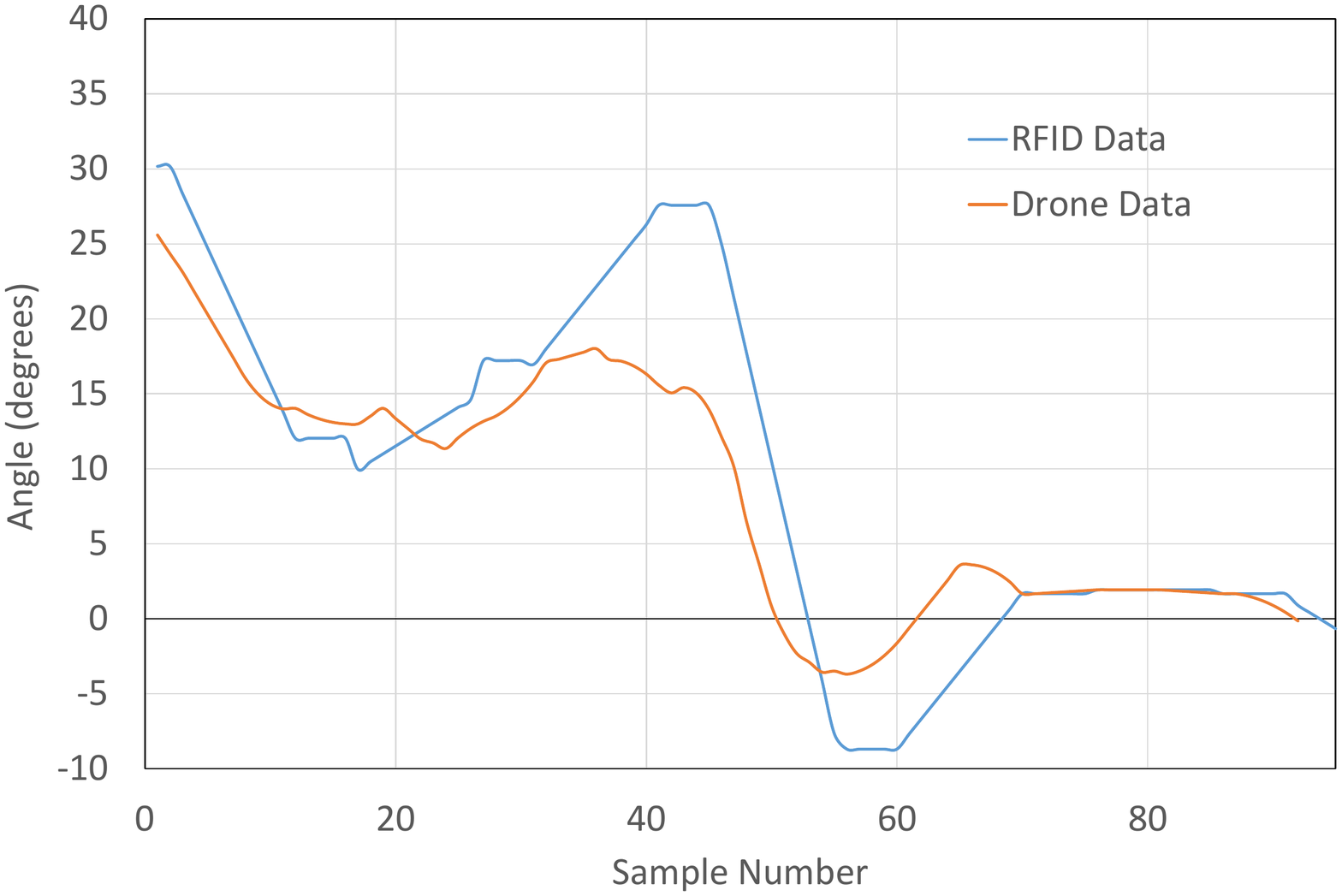}
    \caption{Angle estimation from RFID system and drone's accelerometer: +30 to 0 degrees.}
    \label{anglepath}
\end{figure}

When implemented in a control system, the drone could be rotated to any orientation within the range: $\pm$ 45 degrees. However, it was noted that the accuracy of the system seems to improve the closer the drone's orientation was to zero degrees (i.e. perpendicular to the antenna). Figure \ref{anglepath} displays the orientation as recorded by the RFID reader and by the drone's IMU. A link to a video of this in operation can be found in Section \ref{Demonstrations}.

\begin{table}[h!]
\caption{RFID-based angle estimation: average estimation error in recorded estimations}
\centering
\begin{tabular}{|
>{\columncolor[HTML]{EFEFEF}}c |c|c|c|c|c|}
\hline
True Angle ($^{\circ}$)         & 0       & -30      & 30      & -15      & 15      \\ \hline
Avg Angle Found ($^{\circ}$) & 0.6570  & -30.2382 & 30.7499 & -16.0704 & 15.6625 \\ \hline
Error ($^{\circ}$)              & -0.6570 & 0.2382   & -0.7499 & 1.0704   & -0.6625 \\ \hline
S. Deviation ($^{\circ}$) & 1.1135  & 1.2762   & 1.0035  & 0.8030   & 0.6567  \\ \hline
\end{tabular}
\vspace{0.1cm}
\label{angletable}
\end{table}

\subsubsection{Control Systems and Applications}

With the information collected and proof that there is enough accuracy to recognise both distance and angle, it is clear that passive RFID tags can be used as zero power sensors in highly critical areas, alleviating fears of loss of power. From this, ideas of personal and airborne navigation through adding tags to moving agents are both feasible and accurate.
A system has already been developed using the results found in this paper, in which the tags alone can distinguish at which angle and position an agent is located, and offer navigation assistance to direct that agent. As discussed above, we imagine that this could be used to aid road crossing with auditory or vibratory feedback given through a user's smartphone or the traffic light system itself. We further imagine that the sequence of traffic lights could be altered to ensure safety of the pedestrian.

\begin{figure}[h]
    \centering
    \includegraphics[scale=0.27]{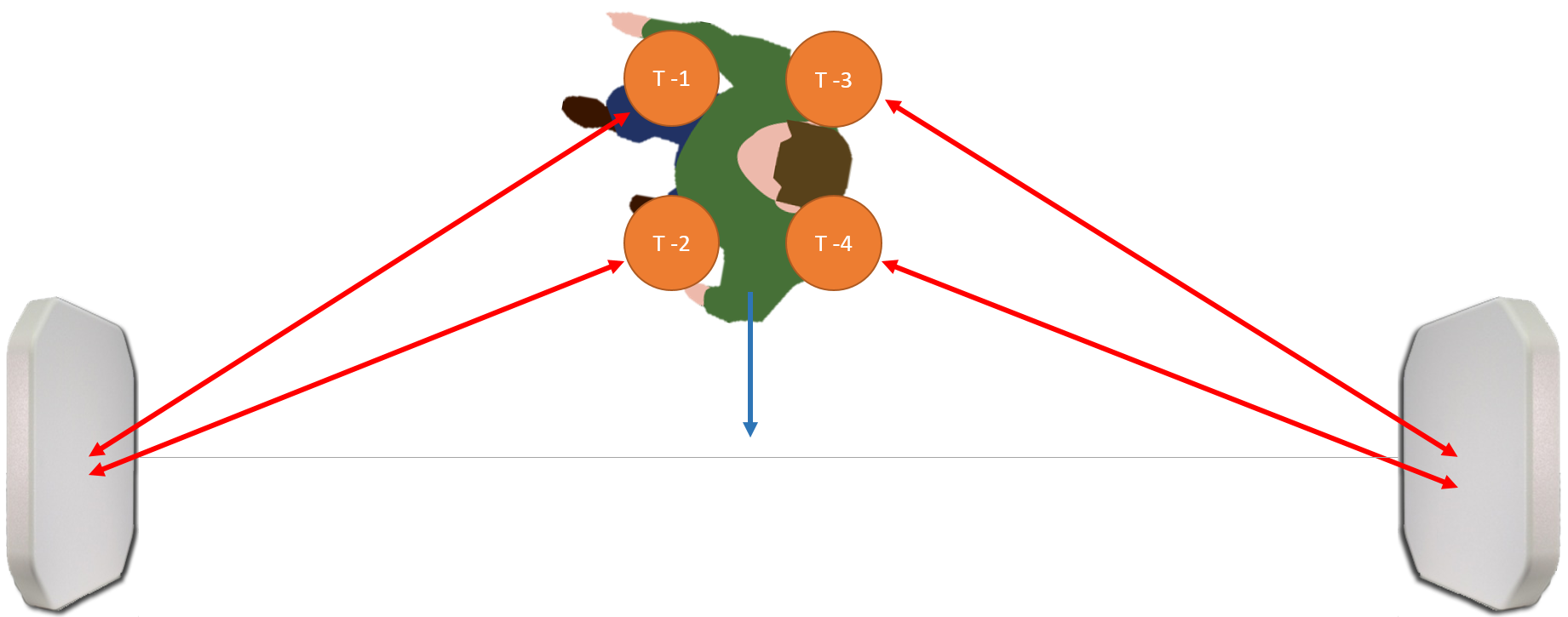}
    \caption{Example of RFID setup for feedback control of a moving agent.}
    \label{TagSetup}
\end{figure}

Using the difference in the RSSI values, it has been found that an agent can be directed to a centre line between 2 antennas, and from there the agent can be accurately maneuvered along such an straight line, as illustrated in Figure \ref{TagSetup}.
Furthermore, using RSSI to estimate the distance of an agent from an antenna accurately in a straight line means that an agent can be controlled and moved/directed to a set location. This has been achieved in airborne tests using a Parrot AR Drone 2.0 and AutoFlight \footnote{{https://electronics.kitchen/autoflight}} drone control software.  

\section{Demonstrations} \label{Demonstrations}

We have prepared a set of movies to demonstrate the key salient points of our results.

{\bf Movie A (Indoor Waypoint Navigation):} Video demonstrating indoor waypoint tracking to a tolerance of 25\,cm. Position is estimated by using 4 Pozyx anchors (UWB) to localize one tag located on the drone.

Link: \texttt{\url{https://youtu.be/pnRgJtCT\_Uw}}

{\bf Movie B (Location Control of Drone using RFID):} Video demonstrating control of drone's location in one dimension to a tolerance of 10\,cm, using RSSI readings from two tags, from two antennas.

Link: \texttt{\url{https://youtu.be/cm7VHIjtH\_U}}

{\bf Movie C (Orientation Control of Drone using RFID):} Video demonstrating control of drone's orientation to a tolerance of 3 degrees, using RSSI readings from two tags, from one antenna. 

Link: \texttt{\url{https://youtu.be/bQjGTuhpIBQ}}

\section{Conclusions}

This paper explores the potential to use a number of technologies in a new and complementary way, to redefine how we look at possible infrastructures and implementations to provide city-wide navigation and control. There is huge scope for extension to this work, in refinement and in developing practical working systems that could be put into operation in minimal time. Using both UWB and RFID technologies, allows for a robust and reliable system that reduces the risk in the case of low battery or power shortages situations, maintaining safety at all times. Overall the work completed here has great potential for improving the daily lives of commuters and residents through the implementation of smart services within the city of the future.

\bibliographystyle{ieeetr}
\bibliography{References}

\end{document}